%% file: YuShi.tex
\documentclass[conference]{IEEEtran}
\usepackage{algorithmic}
\usepackage{algorithm}
\usepackage{mathrsfs}
\IEEEoverridecommandlockouts
\makeatletter
\def\ps@headings{%
\def\@oddhead{\mbox{}\scriptsize\rightmark \hfil \thepage}%
\def\@evenhead{\scriptsize\thepage \hfil \leftmark\mbox{}}%
\def\@oddfoot{}%
\def\@evenfoot{}}
\makeatother 
\usepackage{amsmath,graphicx,amssymb,stfloats,subfigure,multicol}
\usepackage{epsfig,epsf,psfrag,amssymb,amsfonts,latexsym,graphicx,mathrsfs,subfigure}
\usepackage{bbm}
\usepackage{chngpage}
\usepackage[dvips]{color}
\usepackage{textcomp}
\usepackage{hyperref}
\usepackage[hyphenbreaks]{breakurl}

\input ZYmacros

\begin{document}

\title{ An Extended Kalman Filter Enhanced Hilbert-Huang Transform in Oscillation Detection}
\author{\IEEEauthorblockN{Zhe Yu\IEEEauthorrefmark{1},
Di Shi\IEEEauthorrefmark{1},
Haifeng Li\IEEEauthorrefmark{2},
Yishen Wang\IEEEauthorrefmark{1},
Zhehan Yi\IEEEauthorrefmark{1},
Zhiwei Wang\IEEEauthorrefmark{1} }
\IEEEauthorblockA{\IEEEauthorrefmark{1}GEIRI North America\\
\IEEEauthorrefmark{2}State Grid Jiangsu Electric Power Company\\
zhe.yu@geirina.net}
\thanks{This work was funded by SGCC Science and Technology Program under contract No. 5455HJ160007.}}

\maketitle

\begin{abstract}
Hilbert-Huang transform (HHT) has drawn great attention in power system analysis due to its capability to deal with dynamic signal and provide instantaneous characteristics such as frequency, damping, and amplitudes. However, its shortcomings, including mode mixing and end effects, are as significant as its advantages. A preliminary result of an extended Kalman filter (EKF) method to enhance HHT and hopefully to overcome these disadvantages is presented in this paper. The proposal first removes dynamic DC components in signals using empirical mode decomposition. Then an EKF model is applied to extract instant coefficients. Numerical results using simulated and real-world low-frequency oscillation data suggest the proposal can help to overcome the mode mixing and end effects with a properly chosen number of modes.
\end{abstract}


\section{Introduction}
\input intro

\section{Review of Hilbert-Huang transform} \label{sec:II}
\input HHT

\section{An Extended Kalman Filter enhancement}\label{sec:III}
\input EKF

\section{Numerical results}\label{sec:IV}
\input simulation

\section{Conclusion}\label{sec:V}
\input conclusion


{
\bibliographystyle{ieeetr}
\bibliography{Bibs/Journal,Bibs/Conf,Bibs/Book,Bibs/Misc}
}


\end{document}

%% file: ZYmacros.tex
\def\beq{\begin{equation}}
\def\eeq{\end{equation}}
\def\bea{\begin{eqnarray}}
\def\eea{\end{eqnarray}}
\def\ba{\begin{array}}
\def\ea{\end{array}}
\def\bitem{\begin{itemize}}
\def\eitem{\end{itemize}}
\def\benum{\begin{enumerate}}
\def\eenum{\end{enumerate}}

\def\edoc{\end{document}}

\def\eg{{\it e.g., \/}}

\definecolor{bgrd}{rgb}{1,1,1}
\definecolor{grey}{rgb}{0.9,0.9,0.6}
\definecolor{gray}{rgb}{0.5,0.5,0.5}
\definecolor{dkr}{rgb}{0.6,0.2,0.2}
\definecolor{dkg}{rgb}{0,0.5,0}
\definecolor{dkb}{rgb}{0.0,0.1,0.7}
\definecolor{light-gray}{gray}{0.85}



%% file: intro.tex
\IEEEPARstart{P}{ower} systems have encountered a growing number of abrupt and violent dynamics due to the increasing of devices with fast changes, \eg renewable energy \cite{Yi&Dong&Etemadi:17TSG}, distributed generators \cite{Di&Etal:17TSG}, energy storage \cite{Wang&Etal:17PESGM}.  This trend makes new technologies with higher resolutions and capability of dealing with non-stationary systems essential. Wide-area measurement system (WAMS) technology using phasor measurement units (PMUs) provides a possibility to monitor fluctuations in power grids and supply services, \eg low-frequency oscillation detection, that conventional supervisory control and data acquisition (SCADA) systems are not capable of.

With the growth in size of interconnected power systems and participation of unsynchronized distributed energy resources, the phenomenon of oscillation becomes widespread and common, sometimes with non-stationary vibrations. Insufficient damped oscillations reduce the system margin and increase the risk of instability and cascading failure. Fast and precise detection makes it possible to take timely response. Accurate estimation of mode coefficients provides vital information to identify event categories and locations. However, the rapid and dramatic dynamics of power systems make it challenging to update the oscillation information reliably.

There is extensive literature on oscillation detection and estimation. Some of the well-known methods include matrix-pencil method (MP) \cite{Sarkar&Pereira:1995APM,Bounou&Lefebvre&Malhame:1992TPS}, eigenvalue realization algorithm (ERA) \cite{Juang&Pappa:1985JGCD,Peterson:1995JGCD}, Hankel total least-squares (HTLS) \cite{Sanchez&Chow:1997TPS},  Prony methods \cite{Hauer&Demeure&Scharf:1990TPS,Trudnowski&Johnson&Hauer:1999TPS}, and extended Kalman filter \cite{Yazdanian&Etal:2015TPS, Peng&Nair:2012TPS, Jiang&Matei&Baras:2010WSCS}. Most of these methods assume that the oscillation signal is stationary, which is not practical in general. Dealing with real-world PMU data, these methods often suffer from time-variant frequencies, sudden jumps in measurements, and window size selections, which limit their online applications.

Hilbert-Huang transform (HHT) has been applied to power quality analysis with good performance \cite{Senroy&Etal:07TPS}. It was first proposed in \cite{Huang&Etal:98RSL} and applied to oscillation analysis in power systems by \cite{Ruiz&Etal:2005PSCC}. The capability of dealing with non-stationary and nonlinear systems makes it suitable for fast changing signals. However, there are several significant drawbacks of HHT. First, it can not separate closely located frequencies well, while in power systems, frequencies of oscillations typically distribute in a narrow band\cite{Kundur:book94}. Second, the end effects of empirical mode decomposition (EMD) makes the results on both ends of the data set meaningless. Last but not least, HHT is based on local characteristics which makes it sensitive to not only signal dynamics but also noises.

\subsection{Summary of results}
To overcome the disadvantages of HHT, an extended Kalman filter (EKF) approach is combined with HHT. Hilbert-Huang transform is first applied to oscillation signals to remove the possible dynamic DC components. Then the sum of IMFs are fed to EKF to extract coefficients of oscillations. The oscillation signal is first formulated in a nonlinear model with instantaneous amplitude, frequency, and damping as the system states. The initial estimate and order of model can be determined by the HHT results. Given these coefficients, the instantaneous frequencies and damping factors can be directly estimated. The performance of proposal is examined by non-stationary data sets.

%% file: HHT.tex
The Hilbert-Huang transform \cite{Huang&Etal:98RSL} combines the empirical mode decomposition and the Hilbert spectral analysis. It first breaks down signals into various intrinsic mode functions (IMFs), which forms a complete and near-orthogonal basis for the original signals.  Then Hilbert spectral analysis is applied to each IMF to extract instantaneous frequency. HHT is adaptive and highly efficient dealing with time series. The decomposition is based on the local characteristics of data thus it is suitable for nonlinear and non-stationary processes.
The standard EMD process is summarized in Algorithm \ref{alg:EMD}.
\begin{algorithm}[ht]
\caption{Empirical Mode Decomposition}
\label{alg:EMD}
\begin{algorithmic}
\STATE $1$. Set the original signal as $r_0$ and $i=1$.
\STATE $2$. While the number of extrema in $r_{i-1}$ is larger than 2
\bitem
\item  Set $j=1$ and $h_j=r_{i-1}$.
\item  While $h_j$ does not satisfy IMF criterias \cite{Huang&Etal:98RSL}
\bitem
\item Find successive peaks and valleys of $h_j$.
\item Interpolate peaks and valleys with a cubic spline to form upper and lower envelops $e_u$ and $e_l$
\item Obtain a new estimate $h_{j+1}=h_j-(e_u+e_l)/2$ and $j=j+1$.
\eitem
end
\item Get IMF ${c_i=h_j}$ and residue ${r_i=r_{i-1}-c_i}$. Set ${i=i+1}$.
\eitem
\STATE ~~~end
\STATE $3$ The original signal is decomposed as $r_0=\sum_{k=1}^ic_k+r_i$.
\end{algorithmic}
\end{algorithm}

In each iteration of EMD, interpolation is used to form upper and lower envelopes of the signal. However, during the interpolation, at least one data point outside each end of the data set is required to obtain stable spline. If not chosen properly, the extending data points would cause vibration and distortion in envelopes, referred to as end effects, and make the IMF meaningless.

Variant attempts have been made to improve the end effects in EMD. A linear interpolation is first used to predict the extrema outside the data set, after which a mirroring approach is proposed in \cite{Zhao&Huang:01JZUS}. The mirroring approach reflects the data at both ends and forms envelopes using the reflected signal. Then the truncation of the longed envelop is used as the ones of the original signal. A least square polynomial extending is presented in \cite{Zhang&Zhu&Shen:10CAR}. In \cite{Kokes&Nguyen:11AFEH}, authors proposed a constrained cubic spline to mitigate overshoot phenomenon. However, the estimation error of the extrema beyond the end still troubles HHT.

Another problem of applying HHT in oscillation detection is mode mixing. In each iteration of EMD, only the highest frequency component is designed to be extracted. However, the narrow band of low-frequency oscillation (0.1-2.5Hz) in power systems makes HHT hard to distinguish different frequency components. Multiple frequency components are typically mixed in one IMF, which makes the following Hilbert transform meaningless.

Great efforts have been invested in improving HHT. In \cite{Deering&Kaiser:05ICASSP,Deering:06Dissertation}, authors proposed EMD with masking techniques. It analyzes the original signal using FFT and adds a masking signal using the highest frequency in the original one. Then EMD is carried out to the modified signal, and the highest frequency component is extracted. Similar approaches are presented in \cite{Senroy&Suryanarayanan:07PESGM} and \cite{Laila&Etal:09TPS}. Another approach is ensemble empirical mode decomposition (EEMD) \cite{Wu&Huang:09AADA}. The general idea is to add white noise to the original signal and generate more samples. Then EMD is applied to each new sample, and the (ensemble) means of corresponding IMFs are obtained as the final result. The added white noises cancel each other and significantly reduce the mode mixing. However, these approaches are not designed particularly for power system oscillation detection, where the frequencies locate closely and the dynamic is relatively slow. Thus the performance is far from satisfaction.

%% file: EKF.tex
To overcome the mode mixing and end effects problems of HHT while inheriting the capability of dealing with dynamics in signals, an extended Kalman filter approach is proposed. The extended Kalman filter method of oscillation detection was first proposed in \cite{Yazdanian&Etal:2015TPS} under the stationary assumption. A multi-channel extension was developed in \cite{Yu&Shi:17CSEE}. Here, we extend the algorithm in a dynamic context and examine its performance in instantaneous coefficients estimation. The EKF algorithm provides more smooth and accurate instantaneous estimations and is able to distinguish closely located frequencies. With properly chosen coefficients, the EKF can also provide convergent estimation thus overcomes the end effects. The algorithm is shown in Algorithm \ref{alg:HHT-EKF}.
\begin{algorithm}[ht]
\caption{An Extended Kalman Filter enhanced HHT}
\label{alg:HHT-EKF}
\begin{algorithmic}
\STATE $1$. Apply EMD to the oscillation signal and get IMFs.
\STATE $2$. Sum all IMFs besides the DC component ($<0.1$Hz) to construct a new signal $y=\sum_{k=1}^i c_k$.
\STATE $3$. Determine the number of mode and initial estimate of EKF.
\STATE $4$. Apply the EKF to analyze the new signal $y$.
\end{algorithmic}
\end{algorithm}

In this section, we present how to model an oscillation signal into a system state model and how to use the EKF to estimate the instantaneous frequencies.
\subsection{System State Model}
 Here we consider a discrete system where $y[k]$ represents signal $y$ at the  $k$th time instant. After removing the DC component, the reconstructed signal $y$  from Step 2 of Algorithm \ref{alg:HHT-EKF} can be expressed as follows.
\begin{equation}\label{eqn:signal}
\begin{array}{l}
y[k]=\sum_{l=1}^LA_{l}\exp(-\frac{\sigma_l[k]k}{f_s})\cos(\frac{\omega_l[k]k}{f_s}+\phi_{l})+\varepsilon[k],
\end{array}
\end{equation}
where $L$ is the number of oscillation modes, ${A_{l}\in \mathbb{R}}$ the amplitude of mode $l$,  $\sigma_l[k]$ the damping factor, $\omega_l[k]$ the instantaneous frequency, $\phi_{l}$ the phase angle, $f_s$ the sampling rate, and $\varepsilon[k]$ the measurement error. The measurement noise is assumed to be a white Gaussian noise with zero mean and standard deviation ${R_k}$.

Inspired by \cite{Yazdanian&Etal:2015TPS}, we formulate a  nonlinear system whose states contain frequencies and damping factors of the oscillation modes. Consider a sinusoid signal as follows.
\[
\begin{array}{rl}
s_{l}[k]&\triangleq A_{l}\exp(-\frac{\sigma_l[k]k}{f_s})\cos(\frac{\omega_l[k]k}{f_s}+\phi_{l})\\
&=\exp(-\frac{\sigma_l[k]k}{f_s})A_{l}[\cos(\frac{\omega_l[k]k}{f_s})\cos(\phi_{l})\\
&\mathrel{\phantom{=}}-\sin(\frac{\omega_l[k]k}{f_s})\sin(\phi_{l})]\\
&=\exp(-\frac{\sigma_l[k]k}{f_s})[B^c_{l}\cos(\frac{\omega_l[k]k}{f_s})+B^{s}_{l}\sin(\frac{\omega_l[k]k}{f_s})],
\end{array}
\]
where ${B^c_{l}\triangleq A_{l}\cos(\phi_{l})}$ and ${B^s_{l}\triangleq -A_{l}\sin(\phi_{l})}$.
Consider the evolution of the sinusoid signal as follows.
\[\hspace{-2em}
\begin{array}{l}
s_{l}[k+1]\\
~~=\exp(-\frac{\sigma_l[k+1](k+1)}{f_s})B^c_{l}\cos(\frac{\omega_l[k+1](k+1)}{f_s})\\
~~\mathrel{\phantom{=}}+\exp(-\frac{\sigma_l[k+1](k+1)}{f_s})B^{s}_{l}\sin(\frac{\omega_l[k+1](k+1)}{f_s})\\
~~=\big[B^c_{l}[\cos(\frac{\omega_l[k+1]k}{f_s})\cos(\frac{\omega_l[k+1]}{f_s})-\sin(\frac{\omega_l[k+1]k}{f_s})\sin(\frac{\omega_l[k+1]}{f_s})]\\
~~\mathrel{\phantom{=}}+B^s_{l}[\sin(\frac{\omega_l[k+1]k}{f_s})\cos(\frac{\omega_l[k+1]}{f_s})+\cos(\frac{\omega_l[k+1]k}{f_s})\sin(\frac{\omega_l[k+1]}{f_s})]\big]\\
~~\mathrel{\phantom{=}}\times\exp(-\frac{\sigma_l[k+1]}{f_s})\exp(-\frac{\sigma_l[k+1]k}{f_s}).
\end{array}
\]

Define system states as instantaneous magnitudes, frequencies, and damping factors as follows.
\[\hspace{-1em}
\begin{array}{l}
x_{l}[k]\triangleq\left[
\begin{array}{c}
x^c_{l}[k]\\
x^s_{l}[k]\\
x^\omega_{l}[k]\\
x^\sigma_{l}[k]
\end{array}\right]=\left[
\begin{array}{c}
B^c_{l}\exp(-\sigma_l[k]k/f_s)\cos(\omega_l[k]k/f_s)\\
B^s_{l}\exp(-\sigma_l[k]k/f_s)\sin(\omega_l[k]k/f_s)\\
\omega_l[k]\\
\sigma_l[k]
\end{array}\right]
\end{array}
\]
The state transition is presented as follows.
\begin{equation}\label{eqn:stateTransition}
\begin{array}{cl}
x^c_{l}[k+1]&=\exp(-\frac{x^\sigma_l[k]}{f_s})\cos(\frac{x^\omega_l[k]}{f_s})x^c_{l}[k]\\
&\mathrel{\phantom{=}}-\exp(-\frac{x^\sigma_l[k]}{f_s})\sin(\frac{x^\omega_l[k]}{f_s})x^s_{l}[k]\\
&\mathrel{\phantom{=}}+\epsilon_{l}^c[k],\\
x^s_{l}[k+1]&=\exp(-\frac{x^\sigma_l[k]}{f_s})x^c_{l}[k]\sin(\frac{x^\omega_l[k]}{f_s})\\
&\mathrel{\phantom{=}}+\exp(-\frac{x^\sigma_l[k]}{f_s})x^s_{l}[k]\cos(\frac{x^\omega_l[k]}{f_s})\\
&\mathrel{\phantom{=}}+\epsilon_{l}^s[k],\\
x^\omega_l[k+1]&=x^\omega_l[k]+\epsilon_{l}^\omega[k],\\
x^\sigma_l[k+1]&=x^\sigma_l[k]+\epsilon_{l}^\sigma[k],\\
\end{array}
\end{equation}
where $\epsilon$ is the system noise. Here we assume the frequency and damping of the oscillation change are slow comparing to the sample rate, which is reasonable in power systems.

Define the state of the system as ${x[k]=\big[x_1[k];\cdots;x_L[k]\big]}$ which has a dimension of $4L$-by-1.
We can write the transition in a general form as follows.
\[
x[k+1]=f(x[k])+\epsilon[k],
\]
where the transition function $f(\cdot)$ is nonlinear and can be derived from equation (\ref{eqn:stateTransition}). We assume that $\epsilon[k]$ is a white Gaussian noise with zero mean and covariance matrix $Q_k$.

The measurement equation (\ref{eqn:signal}) can be written as ${y[k]=\sum_{l=1}^L(x^c_{l}[k]+x^s_{l}[k])+\varepsilon[k]}$.
In a more compact form, we obtain the observation function as follows.
\[
y[k]=Hx[k]+\varepsilon[k],
\]
where $H=[1~1~0~0~1~1~0~0\cdots].$
The constructed system is summarized as follows.
\begin{equation}\label{eqn:systemEquation}
\begin{array}{l}
x[k+1]=f(x[k])+\varepsilon[k],\\
y[k]=Hx[k]+\epsilon[k].
\end{array}
\end{equation}

\subsection{Extended Kalman Filter}

Given the system equations (\ref{eqn:systemEquation}), we apply an extended Kalman filter to estimate system states. 
Around the current estimated state, the EKF approximates the nonlinear system by a first-order linearization and  applies a Kalman filter to the linearized system to find the optimal Kalman gain. The nonlinear system model and new measurements are used to calculate new state predictions. This process iterates and the state space model is re-linearized around updated state estimates.

Let $\hat{x}[k|j]$ denote the minimum mean squared error estimate of $x[k]$ given measurements up to and including time $j$ and $P[k|j]$ the covariance matrix of the estimation error. Starting from the initial estimate $\hat{x}[0|-1]$ and $P[0|-1]$, the iteration of the extended Kalman filter for the system equation (\ref{eqn:systemEquation}) is summarized in Algorithm \ref{alg:CEKF}.
\begin{algorithm}[ht]
\caption{Extended Kalman Filter (EKF)}
\label{alg:CEKF}
\begin{algorithmic}
\STATE $1$. Initialize $\hat{x}[0|-1]$ and $P[0|-1]$.
\STATE $2$. For $k=0:N-1$
~~~~\[
\begin{array}{l}
S=R_k+HP[k|k-1]H^T\\
K=P[k|k-1]H^TS^{-1}\\
\hat{x}[k|k]=\hat{x}[k|k-1]+K(y[k]-H\hat{x}[k|k-1])\\
P[k|k]=P[k|k-1]-KHP[k|k-1]\\
\hat{x}[k+1|k]=f(\hat{x}[k|k])\\
P[k+1|k]=F_kP[k|k]F_k^T+Q_k
\end{array}
\]
~~~End
\end{algorithmic}
\end{algorithm}

Here ${F_k=\frac{\partial f(x) }{\partial x}|_{x=\hat{x}[k|k]}}$ is the linearization of the system, and $N$ is the time length of measurements. The prediction process $f(\hat{x}[k|k])$ is stated  as follows.
\[\begin{array}{l}
\hat{x}^c_{l,m}[k+1|k]\\
~~=\exp(-\frac{\hat{\sigma}_l[k|k]}{f_s})[\hat{x}^c_{l,m}[k|k]\cos(\frac{\hat{\omega}_{i}[k|k]}{f_s})-\hat{x}^s_{l,m}[k|k]\sin(\frac{\hat{\omega}_{i}[k|k]}{f_s})],\\
\hat{x}^s_{l,m}[k+1|k]\\
~~=\exp(-\frac{\hat{\sigma}_l[k|k]}{f_s})[\hat{x}^c_{l,m}[k|k]\sin(\frac{\hat{\omega}_{i}[k|k]}{f_s})+\hat{x}^s_{l,m}[k|k]\cos(\frac{\hat{\omega}_{i}[k|k]}{f_s})],\\
\hat{x}^\omega_{l}[k+1|k]=\hat{x}^\omega_l[k|k],\\
\hat{x}^\sigma_{l}[k+1|k]=\hat{x}^\sigma_{l}[k|k].
\end{array}
\]

\subsection{Coefficient Choice}
The accuracy of EKF relies heavily on the choice of initial estimate and order of the system model. In the context of oscillation estimation, a Fast Fourier Transform (FFT) or other similar technology can be employed as a trigger, and the result can be used as a choice of initial points. FFT can estimate the spectra of sinusoids with limited measurements and alarm the operator with potential oscillations if the energy of some frequency differs from noises significantly. These results can be used as inputs to Algorithm \ref{alg:CEKF}, and EKF will estimate the fundamental frequency and damping factors. Another possible choice is to use the results of HHT. The instantaneous frequency and damping results from Hilbert transform can be used as the initial estimate and fed in the EKF.

The proposed EKF algorithm is a model-based method, and its performance relies on the proper choice of coefficients. Tuning of the covariance matrix of noise, $Q_k$ and $R_k$, is the major approach to adjust the performance of EKF. A large $Q_k$ or a small $R_k$ usually causes fluctuation around the actual value, while a small $Q_k$ or a large $R_k$ normally results in poor tracking. In this work, the tuning of coefficients is based on heuristics.

%% file: simulation.tex
In this section, we present numerical results using both simulated and real PMU data collected from real-world system oscillation events. We first apply the proposed algorithms on noisy ring down sinusoid signals and compare the accuracy of the proposed  algorithms with HHT and HHT with masking enhancement \cite{Deering&Kaiser:05ICASSP}. After that, real oscillation data from Jiangsu Electric Power Company in China are examined.

\subsection{Closely Located Frequencies}
In this case, the measurement is two exponentially damped sinusoids with a zero-mean white Gaussian noise as follows.
 \[
y[k]=\sum_{l=1}^2\big[\exp(-\frac{\sigma_l k}{f_s})\cos(\frac{\omega_lk}{f_s}+\phi_l)\big]+5\times\mathbbm{1}(k>\frac{N}{2})+\varepsilon[k],
\]
where the frequencies are ${\omega_1=2\pi}$rad/s and ${\omega_2=3\pi}$rad/s, the damping factors are ${\sigma_1=-0.1}$ and ${\sigma_2=0.01}$, the phase angles are ${\phi_1=0}$ and ${\phi_2=\pi/3}$, $\mathbbm{1}(\cdot)$ is the indicator function, ${f_s=30}$Hz is the sample rates, ${N=150}$ is the data length, and $\epsilon[k]$ is the noise. The measurements contain two closely located frequencies and also a step jump. For static analysis tools, \eg PRONY, the performance is poor due to the dynamics. Thus we compare the EKF enhanced HHT with original HHT \cite{Huang&Etal:98RSL} and HHT with masking enhancement \cite{Deering&Kaiser:05ICASSP}.

\begin{table}
\caption{Closely located frequencies: $Q_k=10^{-9}I$, $R_k=10^{-3}$.}
\centering
\begin{tabular}{|l|c|c|c|c|c|}
  \hline
 Method& MSE $\omega_1$ & MSE $\omega_2$ & Mixing rate  \\
\hline
HHT& $10.00$ & $7.62$  &$85.9\%$\\
\hline
Masking& $7.84$ & $9.00$ & $32.2\%$\\
\hline
EKF& $1.99$ & $2.77$ & $14.7\%$\\
\hline
\end{tabular}
\label{table:staticFreq}
\end{table}
In this simulation, we artificially assume that if the mean of the estimated instantaneous frequency deviates from the real value by $50\%$, the mode is failed to detect. Table \ref{table:staticFreq} summarizes the mean squared errors (MSEs) and the rate of failure of different approaches from 1000 Monte Carlo runs. It can be seen that HHT without enhancement almost always mixes two frequencies. With masking technique, the performance is greatly improved, and EKF enhances the performance furthermore. In those cases that two frequencies are successfully detected, EKF approach also delivers more accurate instantaneous frequency estimate comparing existing methods.

\subsection{Time-variant Frequency}
\begin{figure}[ht]
\centering
\subfigure[Measurements]{
\label{fig:dynamicFreq}
  \includegraphics[width=.5\textwidth]{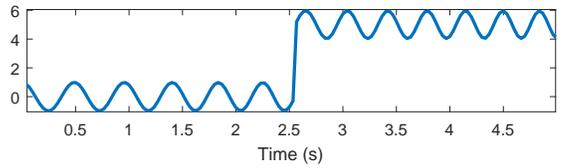}
  }
  \subfigure[Instantaneous frequency estimate]{
\label{fig:freqEstimate}
  \includegraphics[width=.5\textwidth]{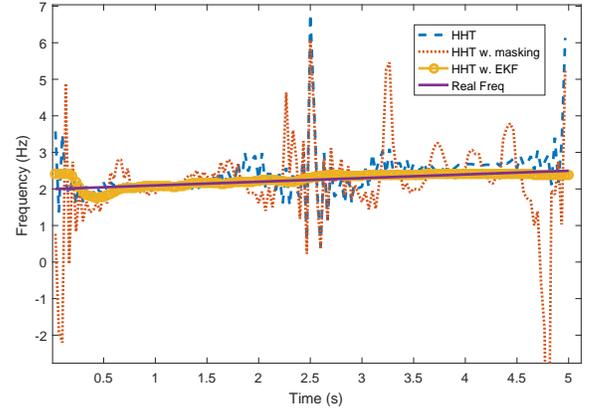}
  }
  \caption{Dynamic frequencies}
\label{fig:testCase2}
\end{figure}

In this case, the measurement is a sinusoid with time-variant frequency and a zero-mean white Gaussian noise as follows.
 \[
y[k]=\cos(\frac{\omega[k]k}{f_s}+\phi)+5\times\mathbbm{1}(k>\frac{N}{2})+\varepsilon[k],
\]
where $\omega[k]=2\pi(1.5+0.5/N\times k)$ is a time-variant frequency. An example of the measurement and the frequency estimate results are shown in Fig. \ref{fig:testCase2}. Table \ref{table:dynamicFreq} summarizes MSEs of different methods from 1000 Monte Carlo runs. To consider the end effects of HHT, we also present MSEs of the middle third of the signal.
 It can be seen that the original HHT and masking enhanced HHT are sensitive to noises and signal changes. On both ends of the signal, the frequency estimate is poor due to the end effects though the time mirroring approach has been employed for mitigation. On the other hand, EKF has a much more stable performance. At the beginning of the signal, the deviation is significant due to the error in initial estimates. However, the frequency estimate rapidly converges to the real value and follows it well all through the process.

It should be noted that EKF has a memory behavior since it takes the past estimates into account. This characteristic makes it slower than HHT when dealing with fast changing signals. Fortunately, in power systems, the change of measurements are relatively slow, and EKF is expected to perform well.

\begin{table}
\caption{Closely located frequencies: $Q_k=10^{-9}I$, $R_k=10^{-3}$.}
\centering
\begin{tabular}{|l|c|c|c|}
  \hline
 Method& MSE $\omega_1$ & MSE $\omega_1$ of the middle third  \\
\hline
HHT& $9.81$ & $6.97$  \\
\hline
Masking& $10.96$ & $7.30$ \\
\hline
EKF& $6.38$ & $3.18$ \\
\hline
\end{tabular}
\label{table:dynamicFreq}
\end{table}
\subsection{Real PMU Data from Jiangsu Electric Power Company}
Jiangsu Electric Power Company, one of the largest provincial power company in China, has installed generation capacity of 100GW and peak load of 92GW. Over 160 PMUs, with thousands of measurement channels, have been installed in the Jiangsu system. These PMUs cover all 500kV substations, a majority of the 220kV substations, major power plants, and all renewable power plants. In this subsection, PMU data collected from real system oscillation events are used to validate the proposal.

\begin{figure}[ht!]
\centering
\subfigure[Real data from Jiangsu]{
\label{fig:realMeasure}
  \includegraphics[width=.22\textwidth]{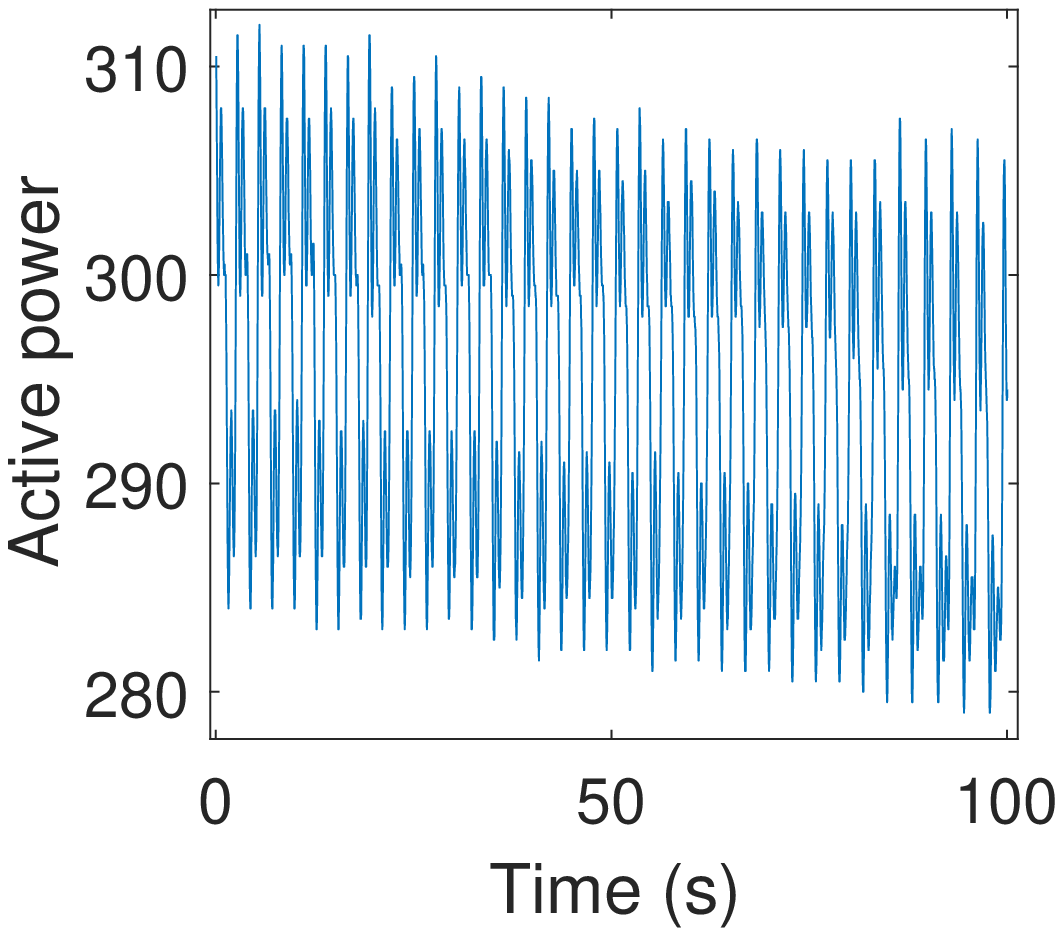}
  }
  \subfigure[Input to EKF]{
\label{fig:imf}
  \includegraphics[width=.22\textwidth]{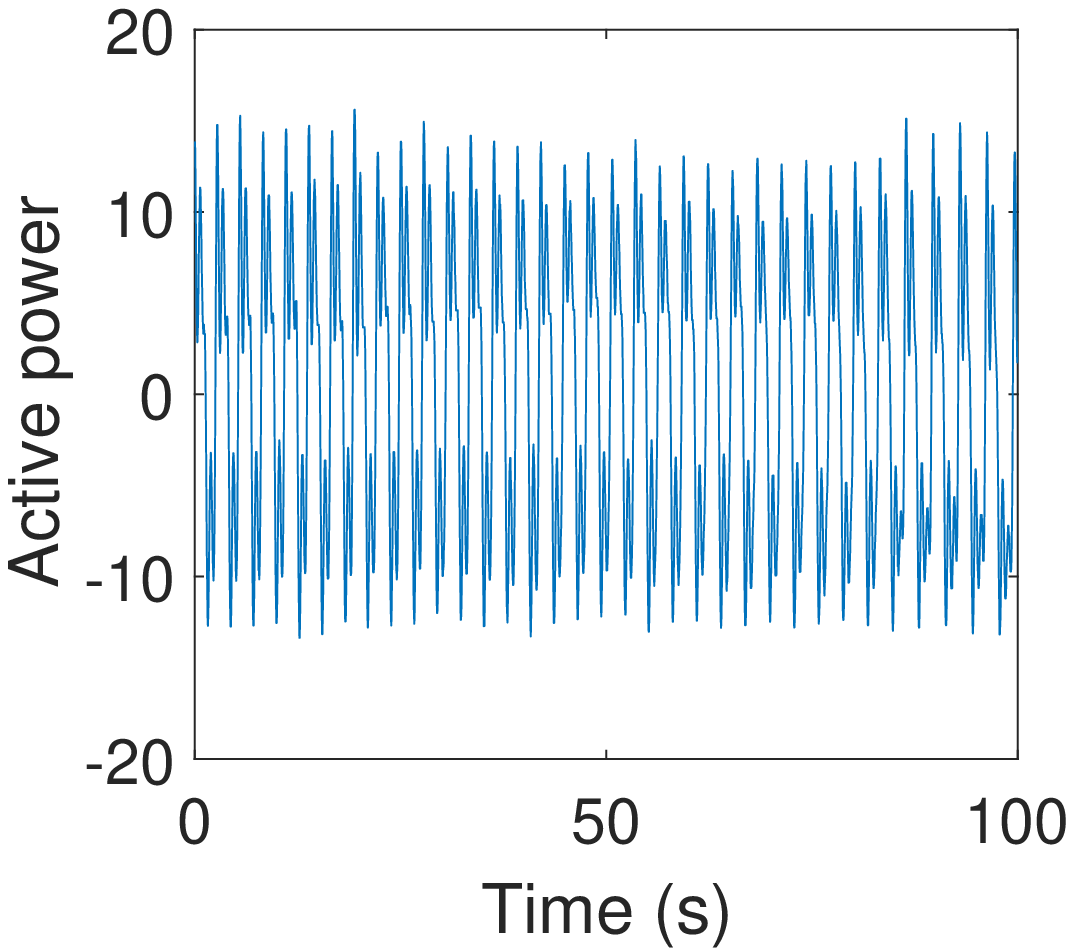}
  }
 \subfigure[Instantaneous frequency estimate]{
\label{fig:EKFresults}
  \includegraphics[width=.5\textwidth]{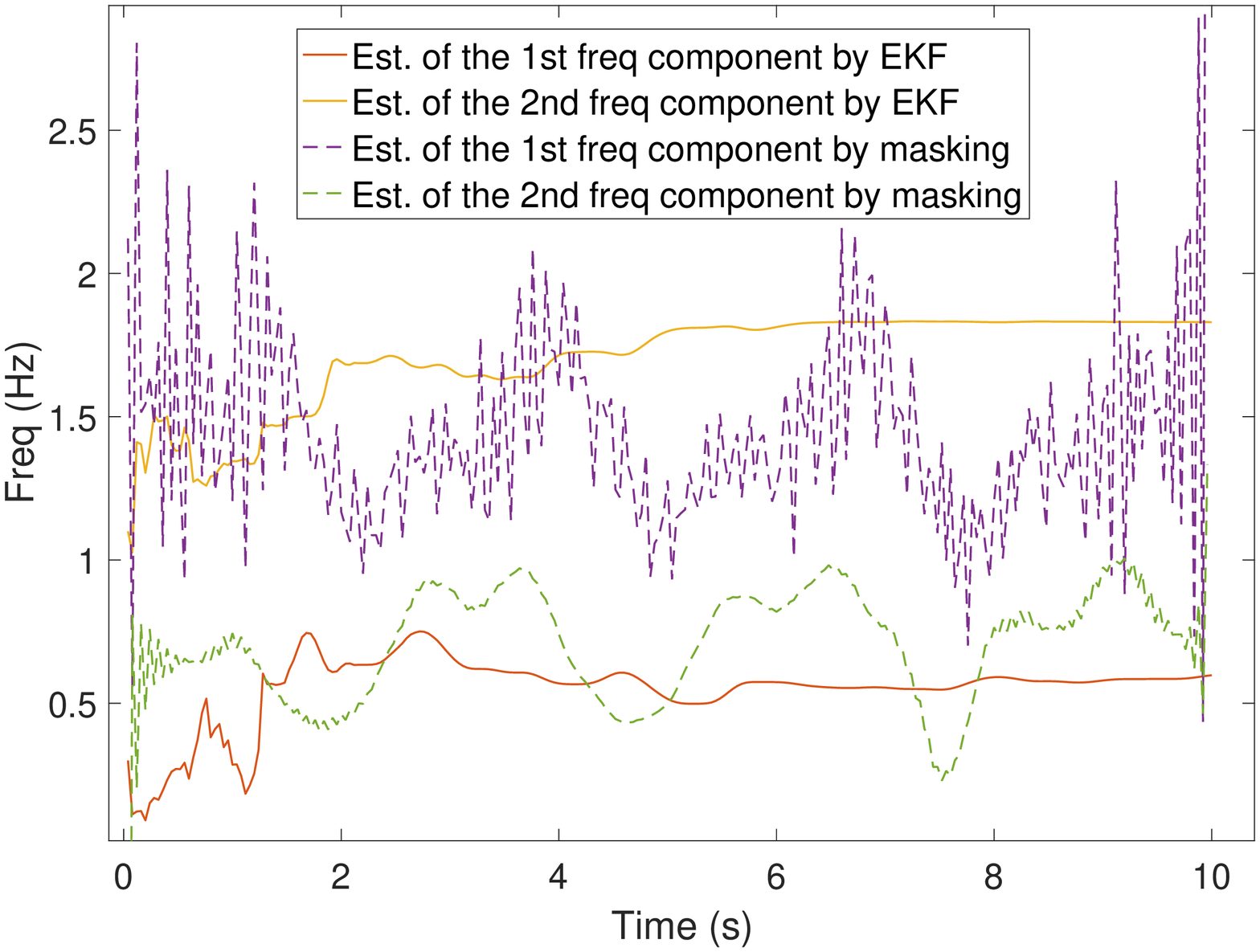}
  }
  \caption{Real PMU measurements from Jiangsu}

\label{fig:testCase3}
\end{figure}

As shown in Fig. \ref{fig:testCase3}, there is a declining trend in the measurements of the active power during the oscillation event. After application of HHT, the trend is removed, and the sum of IMFs is fed to EKF. The EKF results show that there are two frequency components located at $0.5$Hz and $1.5$Hz, which are consistent with the results of masking enhanced HHT. Moreover, the estimate of EKF is more stable than the one of HHT with the masking technique.

%% file: conclusion.tex
An EKF enhancement is proposed to improve the Hilbert-Huang transform. Preliminary results show that the proposal can overcome the mode mixing problem and end effects using simulated data and real PMU measurements. Future work includes how to select the order of mode and initial estimates of EKF automatically so the algorithm can be applied to an online system.